\documentstyle[11pt,twoside,bezier,fancyhea]{article}
\newcommand{\be}{\begin{equation}}
\newcommand{\ee}{\end{equation}}

\newcommand{\bee}{\begin{eqnarray}}
\newcommand{\eee}{\end{eqnarray}}

\title{\Large\bf
EQUILIBRIUM PROPERTIES OF THE
GAS OF ATOMS OF WHICH A PART IS EXCITED
WITHIN CLUSTER EXPANSION METHOD
   }
\author{
   {\sc O.Derzhko, R.Levitskii, O.Chernyavskii}
   \\[1.5ex]
  \it Institute for Condensed Matter Physics
   \\
   \it of the Ukrainian National Academy of Sciences
   \\
   \it 1~Svientsitskii St., UA--290011 Lviv, Ukraine
}
\date{Received February 14, 1995}

\begin{document}
\setcounter{page}{1}
\maketitle

\begin{abstract}
{\small
A gas of atoms some of which are in excited electronic state is under
consideration. Since the lifetime of an excited state is to great extent
larger than the time required for establishing the equilibrium over
translational degrees of freedom the system possesses equilibrium properties.
They are determined essentially by many-particle resonance interactions that
appear in the system of identical differently excited atoms.
The present paper contains the results of numerical calculations
dealing with the examination of virial state equation
and density expansion for
canonical pair spatial distribution function
of the system in question with the accuracy up to
three-particle contributions. The obtained results permit to study the
excited
atoms influence on gas-liquid phase transition and the excimer
molecules formation in the gas with excited atoms.

}
\end{abstract}

The conception of effective interatomic interactions in many-atom
system usually assumes the following arguments. Consider two atoms at
large enough distance so that the operator of interaction energy
between atoms may be taken in the form of multipole series; often one
can restrict himself by dipole-dipole approximation. Making use of
Born-Oppenheimer adiabatic approximation one finds that electron terms
of the system with fixed positions of nuclei play the role of the
potential energy in the nuclear problem and therefore they are the
effective interatomic interaction.  The electron terms of the system in
question may be evaluated with the help of
quantum-mechanical perturbation theory.
Acting according
to the described scheme in the case of two neutral atoms in the ground
electronic state one finds that the first non-vanishing correction
appears within second order of perturbation theory and thus it is
inversely proportional to the six degree of interatomic distance. This is
well known van der Waals interaction. The non-vanishing corrections
arise in the first order of perturbation theory in the case when
identical atoms are in the states with different parity. These
interactions are inversely proportional to the third degree of
interatomic distance and are known as resonance dipole-dipole
interactions.

Although the resonance interactions were well known in condensed matter
physics (e.g. in the theory of molecular excitons) the physical
situation where such interactions may lead to changes in
thermodynamical properties of many-particle system got started to
discuss only in mid-1960s due to V.N.Malnev and S.I.Pekar [1].
Consider a gas of atoms in which due to the external influence (e.g.
irradiation with the frequency that corresponds to excitation energy of
atom) a part of atoms is in the excited electronic state (partially
excited gas (PEG)). The obtained system is essentially nonequilibrium
and relaxes to equilibrium.  The processes of achieving the equilibrium
are characterized by the following times:
\begin{itemize}
\item the mean free path time $\tau_f$ that is connected with
short-range repulsion;
\item the dipole-dipole relaxation time $\tau_d$;
\item the relaxation time
of the processes of exchange by excitation between atoms
$\tau_{exch}$;
\item the time of spontaneous decay of the excited state of an atom
$\tau_{sp}$;
\item the relaxation time of processes of transfer of
electron excitation energy to the energy of translational motion $\tau_c$;
\item the time connected with Doppler width $\tau_D$
\end{itemize}
etc.. It is of special importance, as was noted
by V.N.Malnev and
S.I.Pekar, that {\em {usually the lifetime of
an excited state $(\sim \tau_D, \tau_{sp}, \tau_c \sim 10^{-9} - 10^{-8}s)$
is to great extent larger than
the time that is needed for establishing the equilibrium over
translational degrees of freedom $(\sim \tau_f \sim 10^{-11} - 10^{-10}s)$}}.
This follows both from
experimental data and theoretical estimations [2-14]. Thus, one
faces with the problem of noncomplete equilibrium, i.e. the
equilibrium with respect to translational motion achieves for given
(nonequilibrium) state of electronic subsystem. The equilibrium
properties are expected to be uncommon since they are
determined by resonance interactions that appear in the system
of identical differently excited atoms.

The calculation of equilibrium properties of PEG, however, faces with
some principal difficulties. First, it is necessary to clarify
the meaning of the concept of PEG. Usually [1,15-19] it was supposed
that one has, for example, $N$ two-level atoms, $N_0$ of which are in
the ground electronic state and $N_1$ of which are in excited
electronic state, and $N_0+N_1=N$. However, such definition implies the
possibility to determine the electronic state of each atom, that is
impossible, in a system of interacting atoms. Therefore one
should elucidate whether it is possible to give such "classical"
interpretation to eigenfunctions of Hamiltonian that describes the state
of the system in question in quantum picture.  Considering first the
eigenfunctions of electronic subsystem one easily notes that analytic
with respect to interatomic interaction energy operator eigenfunctions
can be really labeled by the numbers of atoms in ground and excited
states.  Expanding then the eigenfunctions of the total Hamiltonian
in the eigenfunctions of electronic subsystem Hamiltonian one
finds that it is still possible to preserve the numbers of atoms in the
ground and excited states while neglecting by certain effective
interatomic interactions. It seems quite natural since
these effective interactions are responsible for exciting or de-exciting
of atoms due to collisions and hence because of them
one is unable to label Hamiltonian
eigenfunctions by numbers of atoms in the ground and excited electronic
states.
The neglecting by these interactions leads to the
mentioned already limitations
on the time of consideration that should be less than $\tau_c$.
There are even faster processes owing to which the classification
of Hamiltonian eigenfunctions by numbers of atoms in different electronic
states is destroyed and neglecting by which imposes even smaller upper
limit on the observation time
(e.g. the interaction with electromagnetic
field) \cite{20}.

The second problem that one faces with is how to apply Gibbs' scheme for
calculation of equilibrium properties for PEG. Essentially using the
revealed integrals of motion, i.e. the numbers of atoms in the ground
and excited states, one can construct the equilibrium ensembles and find
the distribution for them adopting ergodic hypothesis and proving
Gibbs' theorems or studying the extremal properties of entropy functional.
The most convenient for practical calculations is the grand
thermodynamical ensemble of PEG, that is the ensemble of small
subsystems of PEG that weakly interact with environment via exchange by
energy and atoms in different electronic states. The probability to
detect the system with $N_0$ atoms in the ground state and $N_1$ atoms
in the excited one with energy $E_{j(N_0,N_1)}$ reads
\[
\nonumber
\omega_{j(N_0,N_1)} \sim {\mbox {exp}} \left [\!-\!\beta (E_{j(N_0,N_1)}
\!-\!\mu_0N_0\!-\!\mu_1N_1)
\right]= z_0^{N_0}z_1^{N_1}{\mbox {exp}}(\!-\!\beta E_{j(N_0,N_1)}).
\]
Here $\beta \equiv 1/(kT)$ is the reciprocal temperature and
$z_0 \equiv {\mbox {exp}}(\beta \mu_0)$,
$z_1 \equiv {\mbox {exp}}(\beta \mu_1)$
are the activities of atoms in the ground and excited states.
Note, that in comparison with the grand canonical distribution which
establishes in long runs $\omega_{j(N)} \sim {\mbox {exp}} \left
[-\beta (E_{j(N)}-\mu N) \right]= z^N{\mbox {exp}}(-\beta E_{j(N)})$,
$z \equiv {\mbox {exp}}(\beta \mu )$ is the activity of atoms,
the probability of
states with excited atoms is essentially larger.
Really, consider two possible states of the system, first, with $N_0-1$
atoms in ground state with the energy $E_0$
and $N_1+1$ atoms in excited state with the energy $E_1$,
and second,
with $N_0$ and $N_1$ atoms in the ground and excited states respectively.
The ratio of probabilities of these states
in the latter case is ${\mbox {exp}}\left[ \beta (E_0-E_1)\right]\ll1$
whereas in the former case it is equal to
$[z_1{\mbox {exp}}(-\beta E_1)]/[z_0{\mbox {exp}}(-\beta E_0)]\sim N_1/N_0$.
Hence the states that practically did not
contribute to equilibrium properties appear to be
very significant in PEG \cite{20}.

The calculation of energies of states $E_{j(N_0,N_1)}$
demands some simplifying assumptions.
Assuming the atoms to be two-level ones, neglecting the effects of the
statistics of the atoms, taking interatomic interaction energy operator
in the electrostatic form with further restriction to the dipole
approximation, adopting Born-Oppenheimer approximation and taking for the
solution of a nuclear problem only the zero-order approximation in effective
interatomic interaction, and taking into account the short-range repulsion
by means of introducing the atom radius $\sigma$ one
can write the initial formula for PEG equilibrium properties calculation,
i.e. {\em PEG grand canonical distribution}, in the form:
\begin{eqnarray}
\nonumber
\omega_{j(N_0,N_1)}=\frac{1}{(N_0+N_1)!}z_0^{N_0}z_1^{N_1}\frac{1}{V^N}
{\mbox {exp}} \left(-\beta \sum_{j=1}^{N}\frac{\hbar^2{\bf k}_j^2}{2M}
\right) \\
\times
{\mbox {exp}}\left[-\beta E_{N_0,N_1;\eta}^{el}({\bf R}_1,\ldots,{\bf R}_N)
\right]/\Xi(\beta,z_0,z_1,V),
\label{1}
\end{eqnarray}
all $R_{ij}\ge 2\sigma$ otherwise $\omega_{j(N_0,N_1)}=0$.
Here $\hbar {\bf k}_j$ is the momentum of the $j$th atom,
$M$ is the mass of the atom,
$E_{N_0,N_1;\eta}^{el}({\bf R}_1,\ldots,{\bf R}_N)
\equiv N_0E_0+N_1E_1+U_{N_0N_1}^{\eta}(\ldots,R_{ij},\ldots)$
is the $\eta$th eigenvalue of electronic subsystem Hamiltonian of
$N=N_0+N_1$ atoms in the spatial configuration of atoms
${\bf R}_1,\ldots,{\bf R}_N$, $R_{ij} \equiv \mid {\bf R}_i-{\bf R}_j \mid $,
$\Xi(\beta,z_0,z_1,V)$ is the PEG grand partition function.
Note that in contrast to the corresponding
formula for two component mixture
(1)
contains $1/(N_0+N_1)!$ but not
$1/(N_0!N_1!)$ due to the identity of atoms, and
the effective long-range interatomic interaction in the $\eta$th
state in group of $N=N_0+N_1$ atoms
$U_{N_0,N_1}^{\eta}(\ldots,R_{ij},\ldots)$ is essentially
$N$-particle, has rather complicated dependence on space configuration
${\bf R}_1,\ldots,{\bf R}_N$ and can be found explicity only for few atoms
in the group.
For two-atom groups one has
$U_{20}(R_{12})=-\left(\sqrt{1+\alpha_{12}^2}-1\right)(E_1-E_0)
\approx -\frac{1}{2}\alpha_{12}^2(E_1-E_0)$,
$U_{11}^1(R_{12})=-\alpha_{12}(E_1-E_0)$,
$U_{11}^2(R_{12})= \alpha_{12}(E_1-E_0)$,
$U_{02}(R_{12})=\left(\sqrt{1+\alpha_{12}^2}-1\right)(E_1-E_0)
\approx \frac{1}{2}\alpha_{12}^2(E_1-E_0)$,
where $\alpha_{ij}=a/(R_{ij}/\sigma )^3$,
$a\equiv d^2/[\sigma^3(E_1-E_0)]$,
$d$ is the value of the transitional electrical dipole moment between
the ground and the excited states.
$U_{20}(R_{12})$,
$U_{02}(R_{12})$ and
$U_{11}^1(R_{12})$,
$U_{11}^2(R_{12})$
are familiar van der Waals and resonance dipole-dipole interactions.
More complicated
results of examining the resonance dipole-dipole
interactions in groups of three and four two-level atoms can be
found in \cite{21}.

The only method that permits one to obtain grand thermodynamical
potential $\Omega (\beta,z_0,z_1,V)$ or
grand canonical pair spatial distribution
function ${\cal F}_2(R_{12};\beta,z_0,z_1)$ is cluster expansion
method within the frames of which the contributions to the mentioned
quantities that arise from the groups of larger and larger number of atoms
are taken into account successively.
Formally within such approach one should present
$\Omega (\beta,z_0,z_1,V)$ or
${\cal F}_2(R_{12};\beta,z_0,z_1)$ as a series in the activities $z_0$ and
$z_1$. The contribution from the term that is proportional,
for instance, to $z_0^2z_1$ is determined by $(2+1)$-atom cluster integral
that originates from the group of three atoms two of which are in the ground
state and one in the excited state.
In such a version of the cluster expansion approach
many-atom cluster integrals are finite algebraic sums of terms each of
which tends to infinity with
$V\rightarrow \infty$.
Thus, they should be
regularized, that is presented as algebraic sums of finite terms.
Rather lengthy and tedious details of these calculations are
omitted; these can be found in \cite{22}. Final results can be
written in the form
\bee
\nonumber
-\frac{\beta\Omega(\beta,z_0,z_1,V)}{V}=\\
\nonumber
z_0b_{10}(\beta)+z_1b_{01}(\beta)+z_0^2b_{20}(\beta)+z_0z_1b_{11}(\beta)+
z_1^2b_{02}(\beta)+\\
z_0^3b_{30}(\beta)+z_0^2z_1b_{21}(\beta)+
z_0z_1^2b_{12}(\beta)+z_1^3b_{03}(\beta)+\ldots ,
\eee
\bee
\nonumber
{\cal F}_2(R_{12};\beta,z_0,z_1)=\\
\nonumber
z_0^2f_{20}(R_{12};\beta)+z_0z_1f_{11}(R_{12};\beta)+
z_1^2f_{02}(R_{12};\beta)+\\
\nonumber
z_0^3f_{30}(R_{12};\beta)+z_0^2z_1f_{21}(R_{12};\beta)+
z_0z_1^2f_{12}(R_{12};\beta)+z_1^3f_{03}(R_{12};\beta)\\
\nonumber
+\ldots ,
\;\;{\mbox {if}}\;\; R_{12} \ge 2\sigma,\\
{\cal F}_2(R_{12};\beta,z_0,z_1)=0,
\;\;{\mbox {if}}\;\; R_{12} < 2\sigma,
\eee
where cluster integrals are determined by
\begin{eqnarray*}
 & & b_{20}(\beta)=\frac{1}{2}b_{10}^2(\beta)v(3i_{20}-8), \\
 & & b_{11}(\beta)=\frac{1}{2}b_{10}(\beta)b_{01}(\beta)v(3i_{11}-16),\\
 & & b_{02}(\beta)=\frac{1}{2}b_{01}^2(\beta)v(3i_{02}-8), \\
 & & b_{30}(\beta)=\frac{1}{3!}b_{10}^3(\beta)v^2\left(\frac{9}{2}i_{30}-
144i_{20}+\frac{9}{2}\alpha_{20}+162\right), \\
 & & b_{21}(\beta)=\frac{1}{3!}b_{10}^2(\beta)b_{01}(\beta)v^2\left(
\frac{9}{2}i_{21}\!-\!144i_{20}\!-\!144i_{11}\!+\!\frac{9}{2}\alpha_{20}
\!+\!\frac{9}{2}\alpha_{11}\!+\!486\right), \\
 & & b_{12}(\beta)=\frac{1}{3!}b_{10}(\beta)b_{01}^2(\beta)v^2\left(
\frac{9}{2}i_{12}\!-\!144i_{11}\!-\!144i_{02}\!+\!\frac{9}{2}\alpha_{11}
\!+\!\frac{9}{2}\alpha_{02}\!+\!486\right), \\
 & & b_{03}(\beta)=\frac{1}{3!}b_{01}^3(\beta)v^2\left(\frac{9}{2}i_{03}-
144i_{02}+\frac{9}{2}\alpha_{02}+162\right); \\
 & & f_{20}(R_{12};\beta)=b_{10}^2(\beta)j_{20}, \\
 & & f_{11}(R_{12};\beta)=b_{10}(\beta)b_{01}(\beta)j_{11}, \\
 & & f_{02}(R_{12};\beta)=b_{01}^2(\beta)j_{02}, \\
 & & f_{30}(R_{12};\beta)=b_{10}^3(\beta)v\left[\frac{3}{2}j_{30}+
j_{20}(F_{HS}^{(3)}-16)\right], \\
 & & f_{21}(R_{12};\beta)=b_{10}^2(\beta)b_{01}(\beta)v\left[\frac{3}{2}
j_{21}+(j_{20}+j_{11})(F_{HS}^{(3)}-16)\right], \\
 & & f_{12}(R_{12};\beta)=b_{10}(\beta)b_{01}^2(\beta)v\left[\frac{3}{2}
j_{12}+(j_{11}+j_{02})(F_{HS}^{(3)}-16)\right], \\
 & & f_{03}(R_{12};\beta)=b_{01}^3(\beta)v\left[\frac{3}{2}j_{03}+
j_{02}(F_{HS}^{(3)}-16)\right].
\end{eqnarray*}
Here $v=4\pi\sigma^3/3$, dimensionless $i-$, $\alpha-$, $j-$integrals
are determined by long-range effective interatomic interactions,
for example,
\begin{eqnarray*}
 & & i_{20} \equiv \int_2^{\infty}d\rho \rho ^2 \{ {\mbox {exp}}[
-\beta U_{20}(\rho \sigma )]-1\},\\
 & & i_{11} \equiv \int_2^{\infty}d\rho \rho ^2 \left\{ \sum _{\lambda =1}^2
{\mbox {exp}}
[-\beta U_{11}^{\lambda }(\rho \sigma )]-2 \right\},\\
 & & i_{02} \equiv \int_2^{\infty}d\rho \rho ^2 \{ {\mbox {exp}}[
-\beta U_{02}(\rho \sigma )]-1\};\\
 & & j_{20} \equiv  {\mbox {exp}}[\!-\!\beta U_{20}(\rho_{12}\sigma )],\\
 & & j_{11} \equiv
\sum_{\lambda =1}^2{\mbox {exp}}[\!\!-\beta U_{11}(\rho_{12}\sigma )],\\
 & & j_{02} \equiv  {\mbox {exp}}[\!-\!\beta U_{02}(\rho_{12}\sigma )];
\end{eqnarray*}
explicit
expressions
for other integrals
can be found in \cite{22}, $F_{HS}^{(3)}=8-3R_{12}/\sigma+
(R_{12}/\sigma)^3/16$ if $2\sigma\le R_{12}<4\sigma$ and $F_{HS}^{(3)}=0$ if
$4\sigma\le R_{12}$. The obtained results (2), (3) permit to obtain second
and third virial coefficients in the virial state equation
\begin{eqnarray}
\nonumber
 & & p=\frac{1}{\beta}\rho
\left[ 1+\rho B_2(\beta,c_0,c_1)+\rho^2 B_3(\beta,c_0,c_1)+
\ldots \;\right] ,\\
\nonumber
 & & B_2(\beta,c_0,c_1)=4v-\frac{3}{2}v
\left( c_0^2i_{20}+c_0c_1i_{11}
+c_1^2i_{02}\right) , \\
\nonumber
 & & B_3(\beta,c_0,c_1)=10v^2-\\
\nonumber
 & & \;\;v^2 \left[c_0^3\left( \frac{3}{2}i_{30}-9i_{20}^2+
\frac{3}{2}\alpha_{20}\right) \right.+ \\
\nonumber
 & & \;\;c_0^2c_1\left( \frac{3}{2}i_{21}-9i_{20}i_{11}-
\frac{9}{4}i_{11}^2+\frac{3}{2}\alpha_{20}+\frac{3}{2}\alpha_{11}\right) + \\
\nonumber
 & & \;\;c_0c_1^2\left( \frac{3}{2}i_{12}-\frac{9}{4}i_{11}^2-9i_{11}i_{02}
+\frac{3}{2}\alpha_{11}+\frac{3}{2}\alpha_{02}\right) + \\
 & & \;\;\left.c_1^3\left( \frac{3}{2}i_{03}-9i_{02}^2+\frac{3}{2}\alpha_{02}
\right) \right]
\end{eqnarray}
and two- and three-particle contributions to the series in degrees of
density for canonical pair spatial distribution function
\begin{eqnarray}
\nonumber
 & & F_2(R_{12};\beta,c_0,c_1,\rho)=\\
\nonumber
 & & \;\;F_2^{(2)}(R_{12};\beta,c_0,c_1)+
\rho F_2^{(3)}(R_{12};\beta,c_0,c_1)+\ldots ,
\;\;{\mbox {if}}\;\;R_{12} \ge 2\sigma ,\\
\nonumber
 & & F_2(R_{12};\beta,c_0,c_1,\rho)=0,
\;\;{\mbox {if}}\;\;R_{12} < 2\sigma ,\\
\nonumber
 & & F_2^{(2)}(R_{12};\beta,c_0,c_1)=c_0^2j_{20}+c_0c_1j_{11}+
c_1^2j_{02}, \\
\nonumber
 & & F_2^{(3)}(R_{12};\beta,c_0,c_1)=v\left\{ c_0^3\left[\frac{3}{2}j_{30}
+\left( F_{HS}^{(3)}-6i_{20}\right) j_{20}\right]\right.+ \\
\nonumber
 & & \;\;c_0^2c_1\left[\frac{3}{2}j_{21}+\left( F_{HS}^{(3)}-3i_{11}\right)
j_{20}+\left( F_{HS}^{(3)}-3i_{20}-\frac{3}{2}i_{11}\right) j_{11}\right]+ \\
\nonumber
 & & \;\;c_0c_1^2\left[\frac{3}{2}j_{12}+\left( F_{HS}^{(3)}-\frac{3}{2}
i_{11}-3i_{02}\right) j_{11}+
\left( F_{HS}^{(3)}-3i_{11}\right) j_{02}\right]+ \\
 & & \;\;\left.c_1^3\left[\frac{3}{2}j_{03}+\left( F_{HS}^{(3)}-6i_{02}
\right) j_{02}\right] \right\}.
\end{eqnarray}

It should be noted, that the results obtained are in agreement with the
well known expressions for virial coefficients and the contributions to the
expansion in the density of the pair spatial distribution function,
and the developed approach may be viewed as a generalization
of many-particle Mayer functions formalism \cite{23} for the systems with
quantum resonance interactions.

Let's discuss briefly the results of
{\em numerical calculations that is the main purpose of the present paper}.
Within the adopted approximations the atom is characterized by
the mentioned already
dimensionless parameter $a$. The dimensionless temperature,
distance, pressure, density convenient for presentation of numerical
results are determined by
$\tau =1/[\beta (E_1-E_0)]$,
$\rho_{12}=R_{12}/\sigma $,
$\pi = pv/(E_1-E_0)$,
$\eta =\rho v$.
In fig.1 the
temperature dependences of the second and third virial coefficients for
different cocentrations of excited atoms
$c_1=N_1/(N_0+N_1)$ are depicted. In figs.2,3 two- and
three-particle contributions to pair spatial distribution function at
different concentrations of excited atoms and temperatures are shown.
One can easily note that
in high-temperature limit, when the mutual attraction of atoms becomes not
essential, two- and three-particle contributions correspond to the values
of these quantities for a gas of hard spheres:
$B_2(\beta,c_0,c_1)\stackrel{\beta\rightarrow 0}{\rightarrow}4v$,
$B_3(\beta,c_0,c_1)\stackrel{\beta\rightarrow 0}{\rightarrow}10v^2$,
$F_2^{(2)}(R_{12};\beta,c_0,c_1)\stackrel{\beta\rightarrow 0}{\rightarrow}1$,
$F_3^{(2)}(R_{12};\beta,c_0,c_1)\stackrel{\beta\rightarrow 0}{\rightarrow}
vF_{HS}^{(3)}$. While the temperature decreases the role of long-range
interactions increases. Moreover, it is important what is the concentration
of excited atoms. In the absence of excited atoms the long-range interaction
contributions are caused by van der Waals interaction.
In the presence of
excited atoms there apear the contributions of resonance dipole-dipole
interaction; the concentrations
of atoms in the ground and excited states
control the contributions of different
interactions.
It can be shown that the sum of terms that are resonance interactions
in the group of $l=l_0+l_1$ atoms found in the
first order of pertubation theory is equal to zero,
$\sum_{\lambda}U^{\lambda}_{l_0l_1}(...,R_{ij},...)=0$,
%
%
and therefore the equilibrium properties do not contain a contribution
that is linear in resonance interaction. Thus in dipole approximation
one has the contributions caused by the square of resonance dipole-dipole
interactions that are accompained by a factor
$\beta ^2$ in contrast to the contributions from the van der Waals
interactions which are accompained by factor
$\beta$
(compare $i_{20},i_{02}$ and $i_{11}$, $\;$ $j_{20},j_{02}$ and $j_{11}$).
That is why the presence of excited atoms essentially influences the
temperature dependences of second and third virial coefficients
and two- and three-particle contributions to canonical pair spatial
distribution function.

Due to described peculiarities the presence of excited atoms slightly
influences the PEG isoterms at high temperatures and is significant at
low temperatures. It is worthwhile to discuss the behaviour of the PEG
isoterms with the decreasing of temperature.
One can easily see from (4) and fig.1 that at the
presence of excited atoms with decreasing of temperature the pressure
is influenced by two factors:
decreasing of $\frac{1}{\beta}$ in (4)
and increasing of $\rho^2 B_3(\beta,c_0,c_1)$
(the latter essentially depends on $c_1$). Due to the latter
factor it may happen so that for the fixed density the pressure will
increase with decreasing of temperature.
However, for smaller densities the role of the second factor decreases and
the isoterms depend on temperature in a usual way:
for fixed density with decreasing of temperature the pressure decreases.

Combining (5) and the results depicted in figs.2,3 one finds that
canonical pair spatial distribution function in the absence of excited atoms
weakly depends on temperature whereas in the presence of excited atoms due
to the contributions of resonance dipole-dipole interaction the temperature
dependence is very strong. It is known that
short-range hard sphere repulsion leads to increase of
$F_2^{(3)}(R_{12};\beta,c_0,c_1)$ at the range of distances between
$2\sigma $ and $4\sigma $ whereas the contributions of attraction
leads to decrease of $F_2^{(3)}(R_{12};\beta,c_0,c_1)$ \cite{24}.
Since the latter contribution in the presence of excited atoms
strongly depends on temperature the presence of excited atoms is of
great significance for structural properties.

The obtained results qualitatively agree with such
results for systems with short-range repulsion and long-range attraction.
They are most similar to the results for Stockmayer potential
that is used for a gas of polar molecules \cite{25}.

On the base of obtained virial state equation one can construct in a
standard way van der Waals type state equation
\bee
 & \beta p=\rho /(1-\rho b)-
\rho^2\beta a(\beta), \;\; b=4v, \;
a(\beta)=a^{\prime}+a^{\prime\prime}(\beta),
\nonumber\\
 & a^{\prime}\equiv (c_0-c_1)va^2(E_1-E_0)/32,\;
a^{\prime\prime}(\beta)\equiv c_0c_1va^2\beta (E_1-E_0)^2/16.
\eee
It should be noted that the constant $a(\beta)$ that
describes the change in pressure due to long-range interaction
because of temperature dependence of the contribution to the second
virial coefficient of the resonance dipole-dipole interaction depends on
temperature.
PEG  state equation (6) together with Maxwell rule permits one to construct
gas-liquid coexistence curves in the presence of excited atoms (fig.4)
\cite{26}.

Another type of PEG state equations that was first pointed out in
\cite{15} permits to study excimer molecules formation in PEG \cite{27}.
Taking into account in (2) only two-particle contributions,
neglecting by
$b_{20}(\beta )$ and $b_{02}(\beta )$,
and expressing $z_0$, $z_1$ via $\rho_0=\rho c_0$ and $\rho_1=\rho c_1$
one gets
$\beta p=\rho_0+\rho_1-z_0z_1b_{11}(\beta )$,
$z_1=[\rho_0-z_0b_{10}(\beta )]/[z_0b_{11}(\beta )]$,
$z_0=(\rho_0\!-\!\rho_1\!-\!b_{10}(\beta )$
$\times b_{01}(\beta )/b_{11}(\beta )
\pm \{ [\rho_1\!-\!\rho_0\!+\!b_{10}(\beta )b_{01}(\beta )/b_{11}
(\beta )]^2\!+\!
4\rho_0b_{10}(\beta )b_{01}(\beta )/b_{11}(\beta )\}^{1/2})$
$/[2b_{10}(\beta )]$,
$ z_0z_1b_{11}(\beta )=(\rho +b_{10}(\beta )b_{01}(\beta )/b_{11}(\beta )
\mp \{ [\rho_1-\rho_0+b_{10}(\beta )b_{01}(\beta )$
$/b_{11}(\beta )]^2+
4\rho_0b_{10}(\beta )b_{01}(\beta )/b_{11}(\beta )\}^{1/2})/2$
and therefore
\bee
 & & \beta p=\frac{\rho}{2}
-\frac{b_{10}(\beta )b_{01}(\beta )}
{2b_{11}(\beta )}
\nonumber\\
 & & \pm \frac{b_{10}(\beta )b_{01}(\beta )}{2b_{11}(\beta )}
\sqrt{\left[1+\frac{(\rho_1-\rho_0) b_{11}(\beta )}
{b_{10}(\beta )b_{01}(\beta )}
\right] ^2+4\rho _0
\frac{b_{11}(\beta )}{b_{10}(\beta )b_{01}(\beta )}}.
\eee
The upper sign should be taken in (7) since such a choice leads
to ideal state equation for small densities $\rho $.
Rewriting r.h.s. of equation (7) as $(\overline{N}_0+\overline{N}_1+N_m)/V$,
where
$\overline{N}_0=N_0-N_m$,
$\overline{N}_1=N_1-N_m$
are the numbers of free atoms in the ground and excited states,
and $N_m$ is the number of pairs non-excited atom - excited atom,
that are identified with excimer molecules, one finds
for excimer molecules density
$\rho_m \equiv N_m/V$ that
\bee
 & & \rho_m=\frac{\rho}{2}+
\frac{b_{10}(\beta )b_{01}(\beta )}
{2b_{11}(\beta )}
\nonumber\\
 & & -\frac{b_{10}(\beta )b_{01}(\beta )}{2b_{11}(\beta )}
\sqrt{\left[1+
\frac{(\rho_1-\rho_0)b_{11}(\beta )}{b_{10}(\beta )b_{01}(\beta )}
\right] ^2+4\rho _0
\frac{b_{11}(\beta )}{b_{10}(\beta )b_{01}(\beta )}}.
\nonumber
\eee
In fig.5 the dependences of
excimer molecules density on pressure for several temperatures and
concentrations of excited atoms are depicted. The increasing of excimer
molecule density with the increasing of pressure seems quite natural
and does not contradict to experimental data on the dependence of
radiation intensity of excimer laser on the gas pressure \cite{28}.

To summarize, we present results of equilibrium statistical properties
investigation of a gas with excited atoms within cluster expansion method
with the accuracy up to three-atom groups contributions.
Thermodynamical and structural properties are strongly influenced by
resonance interactions the contributions of which to equilibrium statistical
properties have special temperature dependence.
The obtained results may be developed in different directions.
In particular, they permit
to study the influence of excited atoms on gas-liquid phase transition and
to estimate a number of excimer molecules that form in the gas with
excited atoms.

The authors are gratefully acknowledge many helpful discussions
with Academician I.R.Yukhnovskii
and would like to dedicate the present paper to him on his
seventieth birthday.

\end{document}